\documentclass[traditabstract, rnote, preprint]{aa}
\usepackage{epsfig,graphicx,natbib}
\usepackage{longtable}
\usepackage{amssymb}

\def\0218{\hbox{B\,0218+357}}

\def\PKS1830{\hbox{PKS\,1830$-$211}}
\def\B1422{\hbox{B\,1422+231}}
\def\B1608{\hbox{B\,1608+656}}

\begin{document}

\title{Using gravitationally lensed images to investigate the intrinsic AGN variability}
\author{I. Mart\'i-Vidal\inst{1} \and S. Muller\inst{1}
}

\offprints{\email{mivan@chalmers.se}}
\institute{Department of Earth and Space Sciences, Chalmers University of Technology, Onsala Space Observatory, SE-43992 Onsala, Sweden
}

\date{Accepted in A\&A. Final version}
\titlerunning{Studying jet activity with gravitational lensing}
\authorrunning{Mart\'i-Vidal \& Muller}

\abstract{
We discuss about how the relative flux densities among the images of gravitationally-lensed active galactic nuclei, AGN, can be used to study the intrinsic AGN variability with high accuracy. Multi-frequency monitoring observations of resolved gravitational lenses can allow us to detect signals of very weak variability and also provide information about the jet opacity and structure. As an example, we investigate the variability of the flux-density ratio between the two lensed images of the blazar B\,0218+357, using dual-frequency cm-wave observations. Similar to our previously reported submm-wave observations of the lensed blazar \PKS1830, we observe a clear chromatic variability, starting short before an increase in the flux-density of the blazar. The evolution of the flux-density ratios between the blazar images shows a more clear and rich structure than that of the mere lightcurves of each individual image. The accuracy in the ratio measurements is allowing us to see variability episodes in the blazar that are weaker than the natural scatter in the absolute flux-density measurements. 
A simple opacity model in the jet is used to consistently explain the difference between the flux-density-ratio evolution at the two frequencies.
} 

\maketitle

\section{Introduction}

A fraction of all the observed active galactic nuclei (AGN) are relatively powerful radio sources \citep[e.g.][]{mars802,mars80}. The radio emission from AGN is mainly non-thermal, and originates in relativistic outflows (jets) that are generated in the immediate neighbourhood of the AGN's central engine (which is likely to be a super-massive black hole, SMBH) and can propagate up to intergalactic distances. Strong magnetic fields are thought to play an important role in the formation and propagation of the jets, and they are also necessary to produce the non-thermal emission \citep[e.g.][]{mars80,bland}. 

The magnetic field and the particle density in the jet decrease with distance from the AGN central engine, mainly due to the jet opening angle \citep[e.g.][]{mars80,lob98}. As a result, extremely high magnetic fields may be found at arbitrarily short distances to the central engine \citep[e.g.][]{zaman,mar15}. Such high magnetic fields (and particle densities) close to the jet base produce a frequency-dependent synchrotron self-absorption (SSA) in the jet plasma \citep{bland,lob98}, which is directly related to the well-known position shift of the peak intensity with frequency \citep[the {\em core-shift effect}, e.g.][]{kov08,gab09,mar11} and produces a delayed correlation in the flux-density variability at different radio frequencies \citep[e.g.][]{fromm1,fromm2}. 

With a simplified jet model \citep{lob98}, it is possible to use the core shift and the variability correlation of the frequencies to derive information related to the jet structure. For gravitational lenses (i.e. when there are different images of the same AGN, but seen at different times), the relative flux density among images (the so-called flux-density ratio) can also be used to derive information about the jet properties \citep[e.g.][]{mar13}. In these cases, the precision in the flux-density ratios can be several times higher than that in the absolute flux density of each individual image, since the most important observational systematics (e.g. atmospheric opacity and instrumental response) do not affect the flux-density ratio (assuming that all the lensed images fall within the same small field of view). 

The time delay of the images of a gravitationally lensed AGN introduce a marked time variability in the flux-density ratios (provided that the AGN is intrinsically variable). Opacity effects in the jet imprint an additional frequency-dependent signature, with higher radio frequencies probing closer to the base of the jet (i.e. ``reacting'' first, e.g. \citealt{fromm2}). For these reasons, \cite{mar13} emphasized that multi-frequency monitoring of (radio-bright) lensed AGNs might be a powerful method to study their activity with unprecedented precision and sensitivity and might provide strong observational constraints on the jet physics.

\cite{mar13} presented multi-epoch and multi-frequency ALMA data toward the lensed blazar \PKS1830. While the main focus of the observations was the spectroscopy of absorption lines from an intervening $z$=0.89 galaxy \citep{mul14}, the data were also used to probe the variations of the continuum emission of the blazar. Serendipitously, the ALMA observations were performed at the time of a strong $\gamma$-ray flare, monitored by Fermi-LAT (\citealt{abdo}). Within a time span of about two months during the $\gamma$-ray flare, \cite{mar13} reported large variations ($\sim$30\%) of the (submm) flux-density ratio $\Re$ between the two lensed images, while the flux density of the blazar did not increase by more than $\sim$5\%. In addition, the flux-density ratio showed a remarkable chromatic behaviour, implying a chromatic structure in the blazar jet.

\cite{mar13} proposed a simple model of plasmon ejection (or a shock-in-jet propagation, e.g. \citealt{tur00}) that can naturally explain the temporal and chromatic evolution of the mm/submm $\Re$ in \PKS1830. According to this model, the frequency dependence of $\Re$ is related to opacity effects close to the base of the radio jet. In addition, since the time-lag between the $\gamma$-ray and submm flares is short (a few days at most), the authors suggested that both flares are co-spatial, in agreement with the shock-in-jet scenario of $\gamma$-ray emission (\citealt{val96}). 

Following the same idea of \cite{mar13}, we have retrieved dual-frequency radio monitoring data of the lensed blazar \0218, previously published by \cite{big99}, to investigate the variability of the flux-density ratio between its two images during a three-month monitoring. Although the observing frequencies in this case are much lower than those used in \cite{mar13} (meaning that they probe much longer distances to the jet base), we consider that these observations of \0218 are still a very good example to show the advantages of using multi-frequency monitoring of gravitational lenses for a high-sensitivity study of the jet variability.

\section{Modelling the frequency-dependent jet variability}
\label{THEORSec}

We used the simplified jet model as described in \cite{lob98}. Assuming that the emission originates in the conical region of the jet, the SSA opacity, $\tau$, can be written as a smooth function of the observing frequency, $\nu$, and the distance to the jet base, $R$, in the form

\begin{equation}
\tau_\nu \propto (R\nu)^{-2}.
\label{TauEq}
\end{equation}

We now assume that the bulk of the variability observed in \0218 is due to a jet feature that originates in the jet base and propagates downstream. This feature could be a travelling shock or an over-density in the jet plasma, for instance \citep[as in, e.g.][]{fromm1,mar13}. If such a feature propagates at a constant speed through the jet, then  

\begin{equation}
R\propto t, 
\label{RtEq}
\end{equation}

\noindent where $t$ is the age of the jet feature. Using Eqs.\,\ref{TauEq} and \ref{RtEq}, it is possible to derive the flux-density evolution related to the jet feature. 
If $\Delta I_0$ and $\Delta I_1$ are its emission intensities at frequencies $\nu_0$ and $\nu_1$, respectively, then 

\begin{equation}
\frac{\Delta I_0}{\Delta I_1} = \left(\frac{\nu_0}{\nu_1}\right)^\alpha\exp{(\tau_1 - \tau_0)},
\label{FFShift}
\end{equation}

\noindent where $\alpha$ is the spectral index ($I \propto \nu^\alpha$) and $\tau_i \propto (t\nu_i)^{-2}$ (see Eq.\ref{TauEq}). Hence, under these assumptions, it should be possible to fully derive $\Delta I_1$ from $\Delta I_0$, and vice versa, by just knowing the proportionality constant between $(t\nu)^{-2}$ and $\tau_\nu$. Such a constant would encode information about the physical conditions in the jet and about the kinematics of the jet feature that causes the flux-density variability. For example, using Eqs. 2 and 3 in \cite{lob98}, if the jet feature moves at close to the speed of light, then

\begin{equation}
\left[ \frac{\nu}{\mathrm{GHz}} \frac{t(\tau_\nu=1)}{\mathrm{day}} \right] ^2 = 0.722 \times \frac{N_1 \, B_1 \, \delta \, \phi \, \sin{\theta}}{(1-\cos{\theta})^2(1+z)^2},
\label{THEOR}
\end{equation}

\noindent where $z$ is the redshift, $N_1$ and $B_1$ are the particle density and magnetic field (both in cgs) at 1\,pc from the jet base, $\delta$ is the Doppler boosting, $\phi$ is the jet opening angle, $\theta$ is the jet viewing angle, and $t(\tau=1)$ is the time at which the jet feature reaches the $\tau=1$ surface at frequency $\nu$.

For a gravitational lens with which multiple images of the same AGN are observed, the model given by Eq. \ref{FFShift} would also describe the (frequency-dependent) evolution of the flux-density ratios of the lensed images. Each image would correspond to a different intrinsic time in the AGN, which means that we would probe different values of $t$ (Eq.\,\ref{RtEq}) for the same jet feature as it is seen from the different lensed images. Therefore, this would result in different $\Delta I_0/\Delta I_1$ for each image.

The variability of the images of a gravitational lens can also be used to study weak micro-lensing episodes. For micro-lensing, the variability is not correlated among the images, therefore it will not be possible to fit all the flux-density ratios using the same jet-flaring model (e.g. Eq.\,\ref{FFShift}) in all images. Hence, monitoring with a dense time sampling should allow us to decouple the intrinsic jet variability from the micro-lensing events in each lensed image based on the flux-density ratios. Obviously, such a decoupling will be even easier for lenses with more than two images.

\section{Blazar B\,0218+357}

The blazar \0218 (located at $z$=0.944$\pm$0.002, \citealt{coh03}) shares many similarities with \PKS1830. It is lensed by an intervening ($z$=0.68, \citealt{bro93,car93,wik95}) spiral (\citealt{yor05}) galaxy, which forms two bright and compact images (hereafter A and B) separated by $\sim$0.3$''$, with image B embedded in a faint Einstein ring (\citealt{odea92,pat93}). Molecular absorption is detected (\citealt{wik95}) toward the image A (\citealt{men96,mul07}).

To derive the time delay $\Delta t$ between the two images A and B, \cite{big99} have conducted a three-month VLA monitoring of the source at 8.4\,GHz (X band) and 15\,GHz (U band), between 1996 October and 1997 January. They derived $\Delta t$=10.5$\pm$0.4\,days (image A leading). \cite{big99} measured the flux densities of images A and B by fitting a source model consisting of two point sources (images A and B) plus a broad Gaussian component (for the Einstein ring), directly to the visibilities. \cite{barn} have also estimated the time delay from the correlation of time-resolved $\gamma$-ray observations with Fermi-LAT, obtaining a value of 11.46$\pm$0.16\,days, which is compatible with that of \cite{big99}.

\cite{big99} estimated a flux-density ratio A/B of 3.57$\pm$0.01 and 3.73$\pm$0.01 at 8.4 and 15\,GHz, respectively, which are the values that we use in this paper as the flux-density ratios in the quiescent state of the blazar (i.e. the ratios only related to the lens geometry, without effects coming from the blazar variability). Estimates of the flux-density ratio at shorter (mm-wave) wavelengths \citep{mul07} result in higher values ($\sim$4.2), probably due to the free-free absorption screen, the slightly different locations of the emission throughout the jet (which imply different magnifications), and to the different bias coming from the extended emission (i.e. the Einstein ring), which should be smaller at higher frequencies.

\subsection{Radio light curves of the \0218 images}

\cite{big01} analysed the intra-day variability seen in some parts of the radio light curves at 15\,GHz. These authors discarded the role of extrinsic effects, such as scintillation or microlensing, and concluded that the observed variability is intrinsic to the AGN.

\begin{figure}
\centering
\includegraphics[width=9cm]{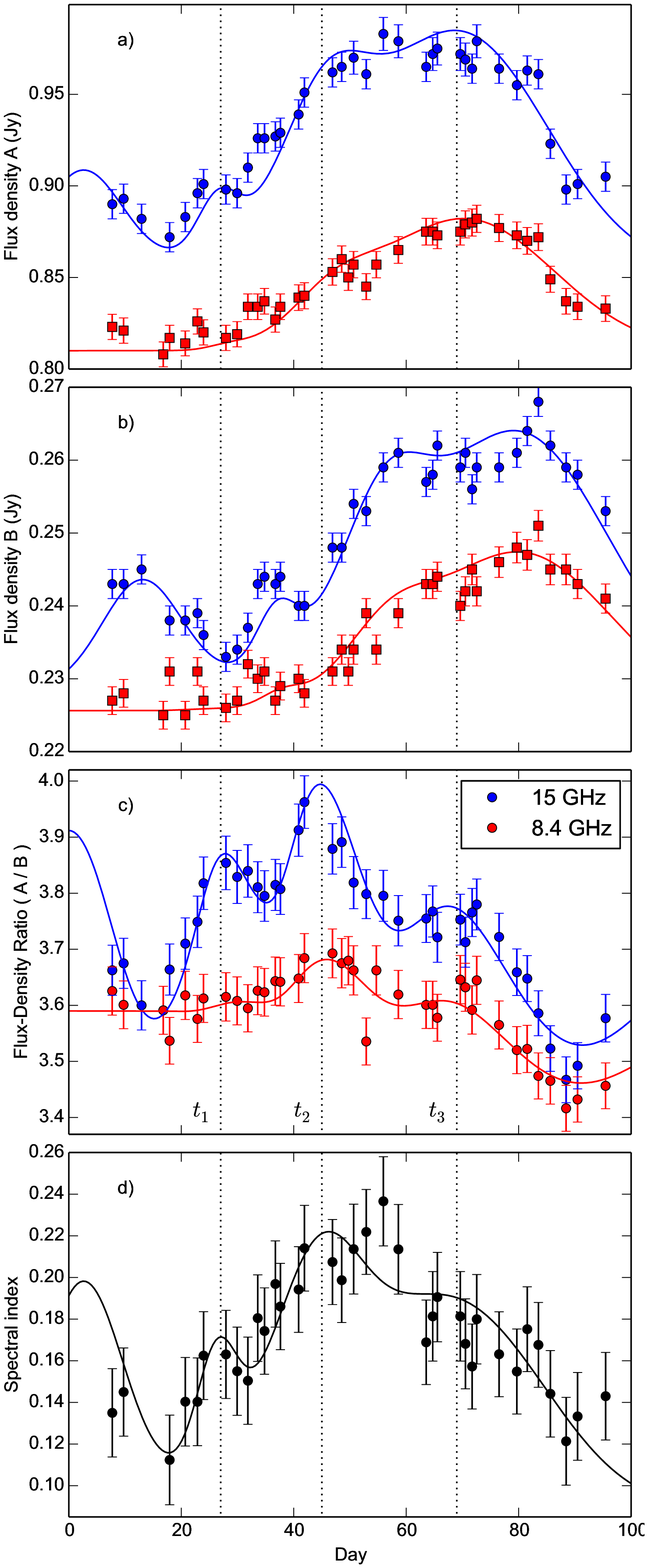}
\caption{From top to bottom: {\bf a)} evolution of the flux density of image A, {\bf b)} evolution of flux-density of image B, {\bf c)} flux-density ratio (A/B), and {\bf d)} spectral index of image A. The dotted lines mark local maxima in the evolution of the flux-density ratio at 15\,GHz. The continuum lines are our model to the data (the 8.4\,GHz model is derived from that at 15\,GHz by an opacity correction with Eq.\,\ref{FFShift}, see text).}
\label{fig1}
\end{figure}

In Fig\,\ref{fig1}\,a we show the light curve of image A during the $\sim$100-day monitoring period (as taken from \citealt{big99}). The evolution of the flux-density ratio (with the leading image, A, in the numerator) is shown in Fig.\,\ref{fig1}\,c (at 8.4\,GHz, $\Re_X$, and at 15\,GHz, $\Re_U$). The spectral index variation of image A is shown in Fig.\,\ref{fig1}\,d. 

The flux ratios are not constant in time, but vary in a range between 3.4 to 3.9, that is, by $\Delta \Re$/$\bar{\Re}\sim$15\%, within the observing period. The evolution of the flux-density ratios is different at the two frequencies, with the flux ratios at 15\,GHz always higher than at 8.4\,GHz, and with more prominent variations. In addition, the flux-density ratio variations do not seem to be always correlated. Regarding the spectral index, it increases (i.e. the spectrum becomes harder) during the first half of the observing campaign, reaching a peak value of $\sim0.23$, and decreases during the second half of the observations to a value of $\sim0.1$. 

We note some small bumps in the light curves, around days 20$-$30 (especially at 8.4\,GHz) and day 80 (8.4\,GHz and 15\,GHz), which do not have any counterpart in the flux-density ratios. The flux-density ratios are unaffected by these bumps because they appear in both images A and B at the same epochs. Hence, these bumps can only be caused by systematics in the absolute flux-density calibration, which contaminate the variability analysis based on total flux densities. When we rely on the flux ratios, however, all these systematics are removed from the analysis.

To facilitate the description, we have identified three epochs, denoted $t_1$, $t_2$, and $t_3$ in Fig.\,\ref{fig1}, corresponding to local maxima in the variations of $\Re_U$. From the beginning of the observing period to $t_1$ ($\sim$26\,days after), $\Re_X$ remains roughly constant at $\sim$3.6, while $\Re_U$ slowly increases from 3.6 to 3.85. The second peak, at $t_2$ ($\sim$20\,days after $t_1$), sees $\Re_U$ increase up to $\sim$3.95, at the same time that $\Re_X$ rises to $\sim$3.7. After $t_2$, the variations of both $\Re_U$ and $\Re_X$ become more tightly correlated, and the ratios eventually converge to a same level of $\sim$3.45. The ratios seem to rise again from the last data points, but as we do not know what happens afterwards, we did not consider these points in the following analysis.

In comparison, the 15\,GHz flux density of image A shows smoother variations, with a net increase from $t_1$ to $t_2$, a plateau roughly between $t_2$ and $t_3$, and a fall at the end of the observing period. The amplitude of the 15\,GHz flux density variations is also smaller than that of the flux ratio (10\% at most). 

Each one of the $t_1$, $t_2$, and $t_3$ episodes (which are too weak to be identified in the light curve) corresponds to a sudden increase in the rate (i.e. the time derivative) at which the flux-density rises in the leading image (i.e. A). These episodes might be related, for instance, to shock-shock interactions 
\citep{fromm1,fromm2}. 

It is clear, thus, that $\Re_U$ is sensitive to (small) changes in the jet emission that cannot be detected in the evolution of the absolute flux density. The $t_1$, $t_2$, and $t_3$ peaks (especially $t_2$ and $t_3$), which are clearly seen in $\Re_U$, are hidden behind the noisy scatter of the absolute flux-density measurements. The variations of the 8.4\,GHz flux density (still of image A) follow the general trend of the variations at 15\,GHz. However, the rise in flux density (between $t_1$ and $t_3$) is more gradual, the peak (near $t_3$) appears to be offset $\sim$10\,days later, and the amplitude of the change is less marked, only $\sim$7\%.

We note that the independent reduction of the same dataset by \cite{coh00} (their Fig.3) shows similar profiles for the variations of $\Re$, although they used a different analysis procedure.

\begin{figure}
\centering
\includegraphics[width=9cm]{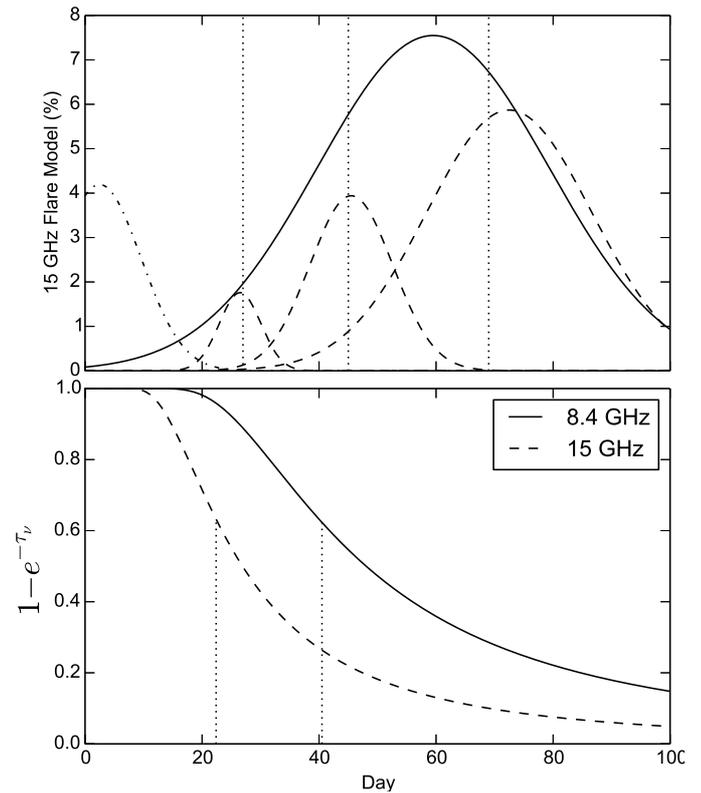}
\caption{Top panel: Gaussian components used to model the variability in the 15\,GHz flux density.  
The dotted lines mark the locations of the local maxima in the flux-density ratio at 15\,GHz. Bottom panel: opacity factor vs. time. The dotted lines mark the epochs where $\tau \sim 1$.}
\label{GaussFig}
\end{figure}

\section{Modelling the flux-density ratios}

We have modelled the flux-density evolution at 15\,GHz using a set of Gaussians, each peaking at a different time (see Fig.\,\ref{GaussFig}, top). We used three Gaussians, roughly centred on $t_1$, $t_2$, and $t_3$ (dashed lines in Fig.\,\ref{GaussFig}, top) and with similar time widths. We also used a fourth Gaussian (continuous line in Fig.\,\ref{GaussFig}, top) with a longer time width to account for the bulk of the longer timescale variability observed in the light curve. A fifth Gaussian (dot-dashed line in Fig.\,\ref{GaussFig}, top) has also been added to model the first four 15\,GHz measurements (this Gaussian has very little effect on the overall light curve). Summing all these Gaussian components, we obtain the model light curve shown in Fig.\,\ref{fig1}\,a (blue line). By applying a time delay of 10.5\,days, we computed the model flux-density of image B (Fig.\,\ref{fig1}\,b, blue line) and the ratios at 15\,GHz that are shown in Fig.\,\ref{fig1}\,c (blue line).

Obviously, the modelling of the 15\,GHz lightcurve (and flux-density ratios) is not unique. We chose a set of Gaussians, some of them approximately centred on the peaks observed in the flux-density ratios, but alternative spline interpolations could have described the flux-density evolution equally well, and using a similar number of free parameters. However, we note that the actual parametrization of the 15\,GHz model light curve does not affect our conclusions about the flux-density relationship between frequencies and/or lensed images. 

Using the model light curve at 15\,GHz and applying Eq.\,\ref{FFShift}, we obtain the model light curve and flux-density ratios at 8.4\,GHz (red lines in Fig.\,\ref{fig1}\,a-c). The flux densities at U and X bands are corrected by

\begin{equation}
\tau = -11.29\times10^4\left(\left[\frac{\nu}{\mathrm{GHz}}\right]\left[\frac{t-t_0}{\mathrm{day}}\right]\right)^{-2}
\label{RelEq}
\end{equation}

\noindent and

\begin{equation}
\left(\frac{\Delta I_U}{\Delta I_X}\right)_{t_0} = 1.12,
\label{RelEq2}
\end{equation}

\noindent with $t_0$ set to 7.7\,days before the first observation. The evolution of the opacity correction at both frequencies is shown in Fig.\,\ref{GaussFig} (bottom). The dotted lines mark the epochs where the opacity is $\sim1$ at U and X bands, which should correspond to the times when the jet feature reaches the jet core at each frequency. Figure \ref{fig1} shows that the light curve (and flux-density ratios) at 8.4\,GHz can be modelled satisfactorily by just using the model at 15\,GHz plus our simple opacity correction. This correction encodes information about the jet structure and 
kinematics (see Eq.\,\ref{THEOR}). 

Assuming that the jet viewing angle falls between 3 and 5 degrees and using reasonable values for the opening angle (1 degree), Doppler boosting (10), and particle density at 1\,pc ($10^3$\,cm$^{-3}$), we estimate a magnetic field between 0.2\,G and 1\,G at 1\,pc, which is in the line of the values given in \cite{lob98} for other AGN (by fitting VLBI observations with this conical-jet model). Our magnetic-field estimate decreases to 0.02$-$0.1\,G for a Doppler boosting of 100.

\subsection{Comparison with $\gamma$-ray modelling}

\cite{barnacka} used the gravitationally lensed $\gamma$-ray variability of \0218 together with a model of the lens mass distribution to estimate the position of the $\gamma$-ray emitting region in the AGN relative to the jet radio cores with milli-arcsecond precision. These authors estimated a separation of $50 \pm 8$\,pc between the radio core and the $\gamma$-ray region, where the $\gamma$-ray region is closer to the central engine. This model is also supported by \cite{spignola}, who did not detect any radio counterpart to the $\gamma$-ray flaring activity in \0218 and interpreted this non-detection as indicative of a full blockage of the $\gamma$-ray region to the radio wavelengths by the SSA. This means that the location of the jet cores at 8.4 and 15\,GHz might be tens of pc away from the central engine, provided that the $\gamma$-ray activity originates close to the central engine. This picture does not contradict our interpretation of the radio emission from \0218, since the self-absorbed radio jet cores can be located at several parsecs ($>10$\,pc at mm-wavelengths) from the central engines, downstream of the jets \citep[e.g.,][]{jorstad,agudo}. Following the standard jet model (Eq.\,\ref{TauEq}), the distance from the central engine to the cm-wave cores could easily increase up to several tens of pc, as was found by \cite{barnacka}.

\section{Conclusions}

The flux-density ratios of the images of gravitationally lensed AGNs have been shown to be remarkably robust observables, especially sensitive to weak AGN variability signals that would otherwise be hidden below the natural scatter of the absolute flux-density measurements.
Future multi-frequency monitoring of radio-bright lensed AGN are expected to provide important clues on the activity in AGN jets (e.g. a new sensitivity window in the study of the statistics of strong or weak flaring events), as well as strong observational constraints on the jet physics, especially at high frequencies (e.g. opacity structure of the jet and effect of stationary shocks).

As an example of the advantages of using gravitational lenses in the study of weak AGN variability, we have investigated the frequency-dependent variation in the flux-density ratio between the two images of the lensed blazar \0218. We used previously published dual-frequency radio monitoring data by \cite{big99}. We showed that the observed flux-density variations can be consistently explained by a simple model of a travelling jet feature plus opacity effects in the blazar jet.


\begin{thebibliography}{}

\bibitem[Abdo et al.(2015)]{abdo} Abdo, A.~A., Ackermann, M., Ajello, M., et al.\ 2015, \apj, 799, 143 
\bibitem[Agudo et al.(2011)]{agudo} Agudo, I., Marscher, A.~P., Jorstad, S.~G., et al.\ 2011, \apjl, 735, L10
\bibitem[Barnacka et al.(2015)]{barnacka} Barnacka, A., Geller, M.~J., Dell'Antonio, I.~P., \& Zitrin, A.\ 2015, arXiv:1511.02891
\bibitem[Blandford \& K\"onigl(1979)]{bland}{Blandford, R.~D. \& K\"onigl, A.\ 1979, \apj, 232, 34}
\bibitem[Biggs et al.(1999)]{big99}{Biggs, A. D., Browne, I. W. A., Helbig, P., et al., 1999, \mnras, 304, 349}
\bibitem[Biggs et al.(2001)]{big01}{Biggs, A. D., Browne, I. W. A., \& Willkinson, P. N. 1999, \mnras, 323, 995}
\bibitem[Browne et al.(1993)]{bro93}{Browne, I. W. A., Patnaik, A. R., Walsh, D., \& Wilkinson, P. N., 1993, \mnras, 263, L32}
\bibitem[Carilli et al.(1993)]{car93}{Carilli, C. L., Rupen, M. P., \& Yanny, B., 1993, \apj, 412, L59}
\bibitem[Cheung et al.(2014)]{barn} Cheung, C.~C., Larsson, S., Scargle, J.~D., et al.\ 2014, \apjl, 782, L14 
\bibitem[Cohen et al.(2003)]{coh03}{Cohen, J. G., Lawrence, C. R., \& Blandford, R. D., 2003, \apj, 583, 67}
\bibitem[Cohen et al.(2000)]{coh00}{Cohen, A. S., Hewitt, J. N., Moore, C. B., \& Haarsma, D. B., 2000, \apj, 545, 578}
\bibitem[Fromm et al.(2011)]{fromm1} Fromm, C.~M., Perucho, M., Ros, E., et al.\ 2011, \aap, 531, A95
\bibitem[Fromm et al.(2015)]{fromm2} Fromm, C.~M., Fuhrmann, L., \& Perucho, M.\ 2015, \aap, 580, A94
\bibitem[Jorstad et al.(2010)]{jorstad} Jorstad, S.~G., Marscher, A.~P., Larionov, V.~M., et al.\ 2010, \apj, 715, 362
\bibitem[Kovalev et al.(2008)]{kov08}{Kovalev, Y.~Y., Lobanov, A.~P., Pushkarev, A.~B., \& Zensus, J.~A.\ 2008, \aap, 483, 759}
\bibitem[Lobanov(1998)]{lob98}{Lobanov A. P. 1998, \aap, 330, 79}
\bibitem[Marscher(1980a)]{mars802} Marscher, A.~P.\ 1980a, \nat, 288, 12
\bibitem[Marscher(1980b)]{mars80}{Marscher A. P. 1980b, ApJ, 235, 386}
\bibitem[Mart{\'{\i}}-Vidal et al.(2011)]{mar11} Mart\'i-Vidal, I., Marcaide, J.~M., Alberdi, A., et al., 2011, \aap, 533, A111 
\bibitem[Mart\'i-Vidal et al.(2013)]{mar13}{Mart\'i-Vidal, I., Muller, S., Combes, F., et al., 2013, \aap, 558, A123}
\bibitem[Mart{\'{\i}}-Vidal et al.(2015)]{mar15} Mart\'i-Vidal, I., Muller, S., Vlemmings, W., et al., 2015, Science, 348, 311 
\bibitem[Menten \& Reid(1996)]{men96}{Menten K.M. \& Reid M.J. 1996, \apj, 465, L99}
\bibitem[Muller et al.(2007)]{mul07}{Muller, S., Gu\'elin, M., Combes, F., \& Wiklind, T., 2007, \aap, 468, 53}
\bibitem[Muller et al.(2014)]{mul14} Muller, S., Combes, F., Gu{\'e}lin, M., et al.\ 2014, \aap, 566, A112
\bibitem[O'Dea et al.(1992)]{odea92}{O'Dea, C. P., Baum, S. A., Stanghellini, C., et al., 1992, \aj, 104, 1320}
\bibitem[O'Sullivan \& Gabuzda(2009)]{gab09} O'Sullivan, S.~P., \& Gabuzda, D.~C.\ 2009, \mnras, 400, 26 
\bibitem[Patnaik et al.(1993)]{pat93}{Patnaik, A. R., Browne, I. W. A., King, L. J., et al., 1993, \mnras, 261, 435}
\bibitem[Spingola et al.(2016)]{spignola} Spingola, C., Dallacasa, D., Orienti, M., et al.\ 2016, \mnras, 457, 2263 
\bibitem[T\"urler et al.(2000)]{tur00} T\"urler, M., Courvoisier, T.~J.~L., \& Paltani, S. 2000, \aap, 361, 850
\bibitem[Valtaoja \& Teraesranta(1996)]{val96}{Valtaoja, E. \& Teraesranta, H., 1996, \aaps, 120, 491}
\bibitem[Wiklind \& Combes(1995)]{wik95}{Wiklind T. \& Combes F. 1995, \aap, 299, 382}
\bibitem[York et al.(2005)]{yor05}{York, T., Jackson, N., Browne, I. W. A., Wucknitz, O., \& Skelton, J. E., 2005, \mnras, 357, 124}
\bibitem[Zamaninasab et al.(2014)]{zaman} Zamaninasab, M., Clausen-Brown, E., Savolainen, T., \& Tchekhovskoy, A., 2014, \nat, 510, 126





\end{thebibliography}
\end{document}